\documentclass[12pt]{iopart}

\usepackage{iopams}

\usepackage{physics}
\usepackage{blkarray, xcolor}
\usepackage{graphicx}
\usepackage{caption}
\usepackage{float}
\usepackage[fencedCode]{markdown}

\usepackage[
    pdfusetitle,
    colorlinks=true,
    linktoc=page,
    allcolors=blue
]{hyperref}

\begin{document}

\title[Simulations identify mechanism mediating non-Arrhenius behavior in LGPS]{Simulations with machine learning potentials identify the ion conduction mechanism mediating non-Arrhenius behavior in LGPS}


\author{ Gavin Winter}
\address{Massachusetts Institute of Technology, 77 Massachusetts Ave, Cambridge, MA 02139}

\author{Rafael G\'{o}mez-Bombarelli }
\address{Massachusetts Institute of Technology, 77 Massachusetts Ave, Cambridge, MA 02139}
\ead{rafagb@mit.edu}

\vspace{10pt}
\begin{indented}
\item[]November 2022
\end{indented}

\begin{abstract}

Li$_{10}$Ge(PS$_6$)$_2$ (LGPS) is a highly concentrated solid electrolyte, in which Coulombic repulsion between neighboring cations is hypothesized as the underlying reason for concerted ion hopping, a mechanism common among superionic conductors such as Li$_7$La$_3$Zr$_2$O$_{12}$ (LLZO) and Li$_{1.3}$Al$_{0.3}$Ti$_{1.7}$(PO$_4$)$_3$ (LATP). While first principles simulations using  molecular dynamics (MD) provide insight into the Li$^+$ transport mechanism, historically, there has been a gap in the temperature ranges studied in simulations and experiments. Here, we used a neural network (NN) potential trained on density functional theory (DFT) simulations, to run up to 40-nanosecond long MD simulations at DFT-like accuracy to characterize the ion conduction mechanisms across a range of temperatures that includes previous simulations and experimental studies. We have confirmed a Li$^+$ sublattice phase transition in LGPS around 400 K, below which the \textit{ab}-plane diffusivity $D^*_{ab}$ is drastically reduced. Concomitant with the sublattice phase transition near 400 K, there is less cation-cation (cross) correlation, as characterized by Haven ratios closer to 1, and the vibrations in the system are more harmonic at lower temperature. Intuitively, at high temperature, the collection of vibrational modes may be sufficient to drive concerted ion hops. However, near room temperature, the vibrational modes available may be insufficient to overcome electrostatic repulsion, thus resulting in less correlated ion motion and comparatively slower ion conduction. Such phenomena of a sublattice phase transition, below which concerted hopping plays a less significant role, may be extended to other highly concentrated solid electrolytes such as LLZO and LATP.

\end{abstract}

%
%
\submitto{J. Phys. Energy}
%
%
%

\section{Introduction}


Materials in the LGPS family exhibit high ionic conductivity, with values between 5 and 20 mS/cm \cite{Kamaya2011AConductor,Kato2020Li10GeP2S12-TypeTransportation} that are on par with  organic liquid electrolytes in commercial lithium-ion batteries. Since a fully occupied Li$^+$ sublattice leads to significant electrostatic repulsion between neighboring cations, the high ionic conductivity of LGPS is thought to arise from inter-cationic repulsion among lithium ions. In particular, this Coulomb repulsion has been postulated originally by He et al. to drive the concerted ion hop that boosts ionic conductivity in these superionic materials systems, namely LGPS, LLZO, and LATP \cite{He2017OriginConductors,Xu2012One-dimensionalConductor}. LGPS features a BCC-like structure with alternating polyhedra of PS$_4$ and GeS$_4$ tetrahedra. Unique to crystalline superionic conducting solids, LGPS exhibits its exceptional ionic conductivity at a stoichiometric composition, unlike other crystalline solid electrolytes, which rely on defects or non-stoichiometry to attain high ionic conductivity \cite{Kato2020Li10GeP2S12-TypeTransportation}.


The temperature dependence of ion conduction in LGPS has been reported experimentally, where non-Arrhenius behavior and temperature-dependent activation energies have been observed.  \cite{AlexanderKuhn2013TetragonalElectrolytes,Weber2016StructuralLi10GeP2S12,Kato2020Li10GeP2S12-TypeTransportation,Kwon2014SynthesisLi10+Ge1+P2-S12}. However, the origin of the change in activation energy remain a matter of debate. Some groups have proposed grain boundary effects to explain non-Arrhenius or shifts in Arrhenius behavior \cite{Kuhn2013Single-crystalLi10GeP2S12}, where grain boundaries may dictate ion transport at low temperatures. However, grain boundary effects have only been shown to be significant below 250 K in LGPS \cite{Kuhn2013Single-crystalLi10GeP2S12}, and at higher temperatures, bulk ion conduction dominates.

Other groups have hypothesized the existence of a continuous (diffuse) phase transition \cite{Weber2016StructuralLi10GeP2S12,Kato2020Li10GeP2S12-TypeTransportation,Kwon2014SynthesisLi10+Ge1+P2-S12,Miwa2021MolecularPotential}, characterized as Type II by Boyce and Huberman \cite{Boyce1979SuperionicDynamics}, only alters the Li$^+$ sublattice. In particular, a transition from quasi one-dimensional ion conduction at low temperature to three-dimensional ion conduction at elevated temperatures has been proposed \cite{Kwon2014SynthesisLi10+Ge1+P2-S12,Kato2020Li10GeP2S12-TypeTransportation,Schweiger2022IonicTransport,Weber2016StructuralLi10GeP2S12,Iwasaki2019WeakLi10GeP2S12,Kuhn2013Single-crystalLi10GeP2S12,Liang2015In-ChannelNMR}. Nevertheless, ionic conductivity measurements are not available for the full range of temperatures over which the transition is thought to occur. Furthermore, in condensed Li$^+$ systems such as LGPS, where the ion conduction mechanism critically relies on Li$^+$-Li$^+$ interactions, the Haven ratio and other metrics that quantify the Li$^+$-Li$^+$ interactions (correlation) have not yet been studied over the range of temperatures from 300 to 800 K to further characterize the non-Arrhenius behavior. The Haven ratio, as detailed in Equation \ref{eqn:haven_ratio}, can be measured experimentally by back-calculating the total diffusivity $D_{\sigma}$ from the ionic conductivity and then comparing that with the tracer diffusivity $D^{tr}$ (equivalent to $D^{*}$). In a review paper, Kato et al. reported total diffusivities $D_{\sigma}$ and calculated Haven ratios at room temperature to be less than 1 \cite{Kato2020Li10GeP2S12-TypeTransportation}, but this all relied on data from a single sample by Kuhn et al., the only working group to report tracer diffusivity $D^*$ in LGPS, as characterized by nuclear magnetic resonance (NMR) \cite{AlexanderKuhn2013TetragonalElectrolytes}. As a result, the values for $D_{\sigma}$ and $D^{*}$ did not come from the same sample of LGPS. Furthermore, Kuhn et al. themselves characterized the room-temperature Haven ratio for LGPS as on the order of 1, by measuring the total diffusivity $D_{\sigma}$ and the tracer diffusivity $D^*$ by NMR for the same sample \cite{AlexanderKuhn2013TetragonalElectrolytes}.


Historically, \textit{ab initio} molecular dynamics (AIMD) has been used to evaluate the diffusivity and ionic conductivity of crystalline electrolytes. These simulations require a costly quantum mechanical calculation of the atomic forces with every femtosecond of simulated time, and consume hundreds of thousands of CPU hours. Consequently, AIMD is constrained in both the length and time scales that can be accessed. Thus, simulation of solid ion conductors with AIMD is limited to high temperatures, where ion hops occur with very high frequency and diffusivity statistics reach convergence with short simulation times on the order of 100 ps. The high-temperature regimes in which diffusivity is evaluated by AIMD often do not overlap with the near-room-temperature regime of proposed commercial application, where diffusivities and ionic conductivities are measured experimentally. Comparing the experiment and simulation requires extrapolating high-temperature AIMD results, under the assumption that there is no non-Arrhenius behavior.

In the case of LGPS, AIMD and NMR have been used to determine the tracer diffusivity (self-diffusivity $D^*$) in the high- ($>$500 K) and low-temperature regimes ($<$500 K), respectively. Kato et al. has acknowledged that no single method has been able to cover the entire range due to the aforementioned limitations of simulations over the temperature ranges with respect to the non-Arrhenius behavior \cite{Kato2020Li10GeP2S12-TypeTransportation}.

Machine learning provides a way to learn a surrogate for the expensive quantum mechanical forces, preserving the accuracy but increasing speed by 4 or more orders of magnitude \cite{Axelrod2022LearningSimulations}. As early as 2017, NN potentials have been used to study ion conduction in crystalline solid electrolytes \cite{Li2017StudyPotential}. Recently, in the composition space of sulfides, NN potential-driven MD was used by Huang et al. \cite{Huang2021DeepConductors}, as well as by Miwa et al. \cite{Miwa2021MolecularPotential}, to study LGPS. Both NNMD studies successfully benchmarked against literature values for high-temperature diffusivity with AIMD. However, the NNMD simulations by Huang et al. overestimated the experimental diffusivity near room temperature, and there were only two datapoints for NNMD simulations near room temperature reported by Miwa et al., which did not allow for a proper Arrhenius fit. These discrepancies between MD simulations and experiment warranted a more comprehensive study of the diffusion and correlation mechanism in LGPS at room temperature. A NN potential is required to drive MD in order to efficiently run simulations on the timescales required to study the effect of correlation on diffusion. Ultimately, since these MD studies are currently limited to bulk ion conduction (i.e. perfect crystals), the effects of grain boundaries cannot be assessed \textemdash the one limitation of this method.

Using NN potentials, which predict the forces on each atom according to the local environment at each timestep, we have been able to access timescales on the order of tens of nanoseconds with MD and achieved quantum chemical accuracy (comparable to \textit{ab initio} methods), allowing us to evaluate diffusivities for each composition of a crystalline material in a faster time than AIMD. This has allowed us to bridge the gap between studies of ion diffusion in the high temperature regime with AIMD and experimental studies of ion diffusion near room temperature.

\section{Methods}
\subsection{Neural Network Interatomic Potential}
The underlying potential used to drive NNMD simulations in this work is a message-passing neural network (MPNN), which initializes atom-wise features and builds a graph within a local neighborhood of interacting atoms that repeatedly exchanges messages through updates of node features. The polarizable atom interaction neural network (PaiNN) is one example of a message-passing NN that has rotationally equivariant representations \cite{Schutt2021} and is the architecture that we employ as a NN potential for driving MD. Unlike rotation-invariant representations, which are limited in their ability to propagate directional information, the rotational equivariant representations of vector features employed by PaiNN allow directional information to be preserved, allowing for higher accuracy NN potentials with less DFT training data. As a message-passing NN, if the cutoff is 5 \AA, and there are 3 convolutional layers (the hyperparameters chosen for this study), due to the updates with message-passing, the actual local environment is (3x5=)15 \AA, and effectively 4-body interactions are learned \cite{Batatia2022ThePotentials}. Intuitively, there is a tradeoff between accuracy and computational speed in considering the cutoff size and how large a local environment one considers an atom/ion interacting with. Upon inference, the total energy can then be predicted for a specific input geometry. On top of the output energy from PaiNN, a repulsive energy term with a $1/r_{ij}^{12}$ dependence similar to the repulsive term accounting for electron cloud overlap in the Lennard-Jones potential, was added onto the output energy from PaiNN in order to prevent collisions of atoms within their own excluded volume and improve the stability of the NN potential. Auto-differentiating the total energy with respect to atomic coordinates yields the negative of the per-atom forces as in Equation \ref{eqn:forces}, where $i$ is the atom index.
    \begin{equation}
    \label{eqn:forces}
        \hat{\vec{F}}_i = - \frac{\partial{\hat{E}}}{\partial{\vec{r}_i}} 
    \end{equation}
In contrast, auto-differentiating the energy with respect to pairwise distances between all atoms allows the calculation of stress as in Equation \ref{eqn:stress}, where $a$ and $b$ are running along $x$, $y$, $z$, and $i$ and $j$ are running along all the atom indices.
    \begin{equation}
    \label{eqn:virial}
        \hat{\vec{f}}_{ij} = \frac{\partial{\hat{E}}}{\partial{\vec{r}_{ij}}}
    \end{equation}
    \begin{equation}
    \label{eqn:stress}
        \hat{\tau}_{ab} = \frac{1}{2 \Omega} \sum_{j=1, i \neq j} (\vec{r}_{ij})_a ( \hat{\vec{f}}_{ij} )_b
    \end{equation}
 Each timestep in MD requires a call to evaluate forces by this method.

\subsection{Generation of training data}

Generating a high-fidelity NN potential for MD requires training data for energy, forces, and stress, corresponding to structures in phase space relevant to the phenomenon being investigated \textemdash in this case ion hopping. Density functional theory (DFT), using the pseudopotential projector-augmented wave method in the Vienna \textit{Ab Initio} Simulation Package (VASP) \cite{Kresse1993AbMetals,Kresse1996EfficientSet,Perdew1996GeneralizedSimple}, was used to generate the training data with: optimizations (relaxations) of periodic structures, nudged elastic band (NEB) calculations, and subsequent single-point calculations of structures. Each datum in the training data consists of a structure (atomic positions and lattice), the energy of that structure, per-atom forces, and a stress tensor as evaluated by DFT. (For details regarding DFT calculations with VASP, see \ref{sec:supporting_dftdetails}. For details regarding dataset availability, see \ref{sec:supporting_data}.) While previous sampling strategies have involved using AIMD as training data \cite{Batzner2022E3-EquivariantPotentials} or random perturbations to atomic positions to generate off-equilibrium structures \cite{Huang2021DeepConductors}, we chose to primarily rely on NEB as a means of creating an initial set of DFT training data that spans the trajectory of structures representative of the ion conduction pathways in a crystal. Using a transition state method such as NEB for sampling helps acquire thousands of off-equilibrium geometries: (1) with greater efficiency per \textit{ab initio} calculation (as compared with sampling by AIMD, where hops may occur once every several thousand timesteps), (2) in physically relevant phase space (unlike sampling by random perturbations).

The equilibrium structure that was used for the NEB calculations was the 50-atom unit cell with stoichiometric LGPS mp-696128 \cite{Jain2013Commentary:Innovation}, and two different concerted ion hops were sampled with NEB. When there are a multitude of possible sites in the crystal framework's void space for the mobile species, tools that employ Voronoi decomposition \cite{Rycroft2009VORO++:C++} or CAVD \cite{He2019CrystalConductors,He2021AConductors} were used to re-populate the interstitial sites with Li, and bond valence site energy methods \cite{Chen2019SoftBVConductors,Wong2021BondSoftBV} provided a guess ion migration path. This starting equilibrium structure has tetragonal symmetry and the zig-zag (Z) arrangement of PS$_4$ / GeS$_4$ polyhedra. Since only the stoichiometric LGPS compound was used in all instances, no vacancy concentration was assumed. Training data exclusively consisted of the 50-atom unit cell and is a sufficient representation for sampling structures from the ion hopping trajectory. However, the MD was all run with a 3,200-atom 4x4x4 supercell to mitigate any finite size effects and provide a wealth of statistics for ion diffusion.

\subsection{Active learning loops}

In order to supplement DFT training data gathered using NEB, a two-pronged active learning approach \cite{Wang2020ActiveLiquids,Ang2021ActiveSurfaces,Axelrod2022LearningSimulations,Schwalbe-Koda2021differentiable} was used to further sample frames from unseen regions in phase space. (1) One approach relied on sampling of frames from an MD trajectory driven by a first-generation neural network (NN) potential. First, non-physical frames from the MD trajectory were filtered out of the pool of structures to sample from. These included: (i) frames with atoms that collided; (ii) frames with a molten framework since they do not represent physically-relevant phase space. Though the Li$^+$ sublattice is expected to be mobile, the criteria for a molten framework was based on the root-mean-squared-displacement for framework atoms (in reference to the initial structure) with respect to a threshold value. Subsequently, the remaining frames in physically-relevant phase space were sampled by their \textit{uncertainty} and by their \textit{diversity} relative to the structures comprising the existing training dataset. To quantify \textit{uncertainty}, the variance in forces was evaluated by an ensemble of NNs. To ensure \textit{diversity} in sampling, low-dimensional embedding of the similarity matrix, quantified by root-mean-squared displacement, between all filtered frames was performed to do k-means clustering. From each of the 4 clusters, 100 frames were selected for singlepoint DFT evaluation of forces and energies. (2) The other approach, developed by Schwalbe-Koda et al. \cite{Schwalbe-Koda2021differentiable}, relied on an ensemble of NNs to identify trajectories that maximize uncertainty (adversarial loss) in the forces or stress, where uncertainty was again measured as the variance in the forces or stress evaluated by the ensemble of NNs. In addition to the variance in forces or stress, the adversarial loss also includes a Boltzmann term according to the structure's predicted energy $\hat{E}$ to determine the frames that were physically accessible in phase space.

\begin{figure}
    \centering
    \includegraphics[width=16cm]{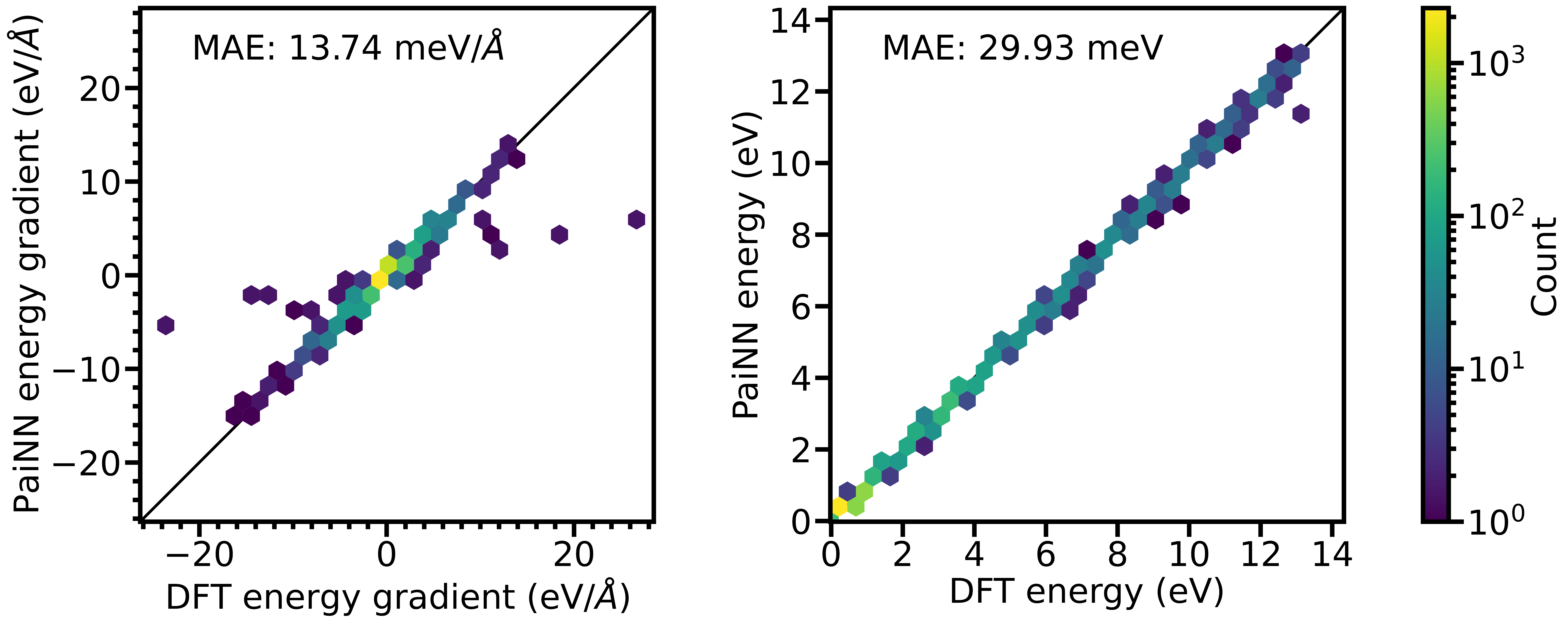}
    \caption{Comparison between PaiNN-predicted properties and DFT-computed properties, as evaluated on the test dataset: (1) energy gradient and (2) total energy. Mean absolute errors (MAE) are below 25 meV/\AA and 50 meV for energies.}
    \label{fig:parity_plot}
\end{figure}

With the collection of more training data through these two approaches and through subsequent generations of training and simulations/sampling, this active learning process continued until the mean average error of the NN on the test set with respect to ground-truth DFT fell below 25 meV/\AA for forces and below 50 meV for energies (as can be seen in Figure \ref{fig:parity_plot}). While the test error is a quantitative metric, nanosecond-long stability is the paramount pre-requisite for MD driven by a NN potential. With this final generation NN potential (details regarding model availability can be found in \ref{sec:supporting_data}), production MD simulations were then run. MD was run on a 3,200-atom 4x4x4 supercell, using a timestep of 1 fs with the Nos\'e-Hoover thermostat in the NVT ensemble and fixed lattice parameters from relaxation at 0 K, as implemented in the atomic simulation environment (ASE) \cite{HjorthLarsen2017TheAtoms} and adapted for NN potentials.

\section{Results}

\subsection{Sublattice phase transition and distinct ion conduction mechanisms in LGPS: bridging the gap between computation and experiment}

After running an MD simulation at each of the fixed temperatures with the NVT ensemble, the self-diffusivity $D^*$ was evaluated using Equation \ref{eqn:self_diffusivity}.
\begin{figure}
    \centering
    \includegraphics[height=18cm]{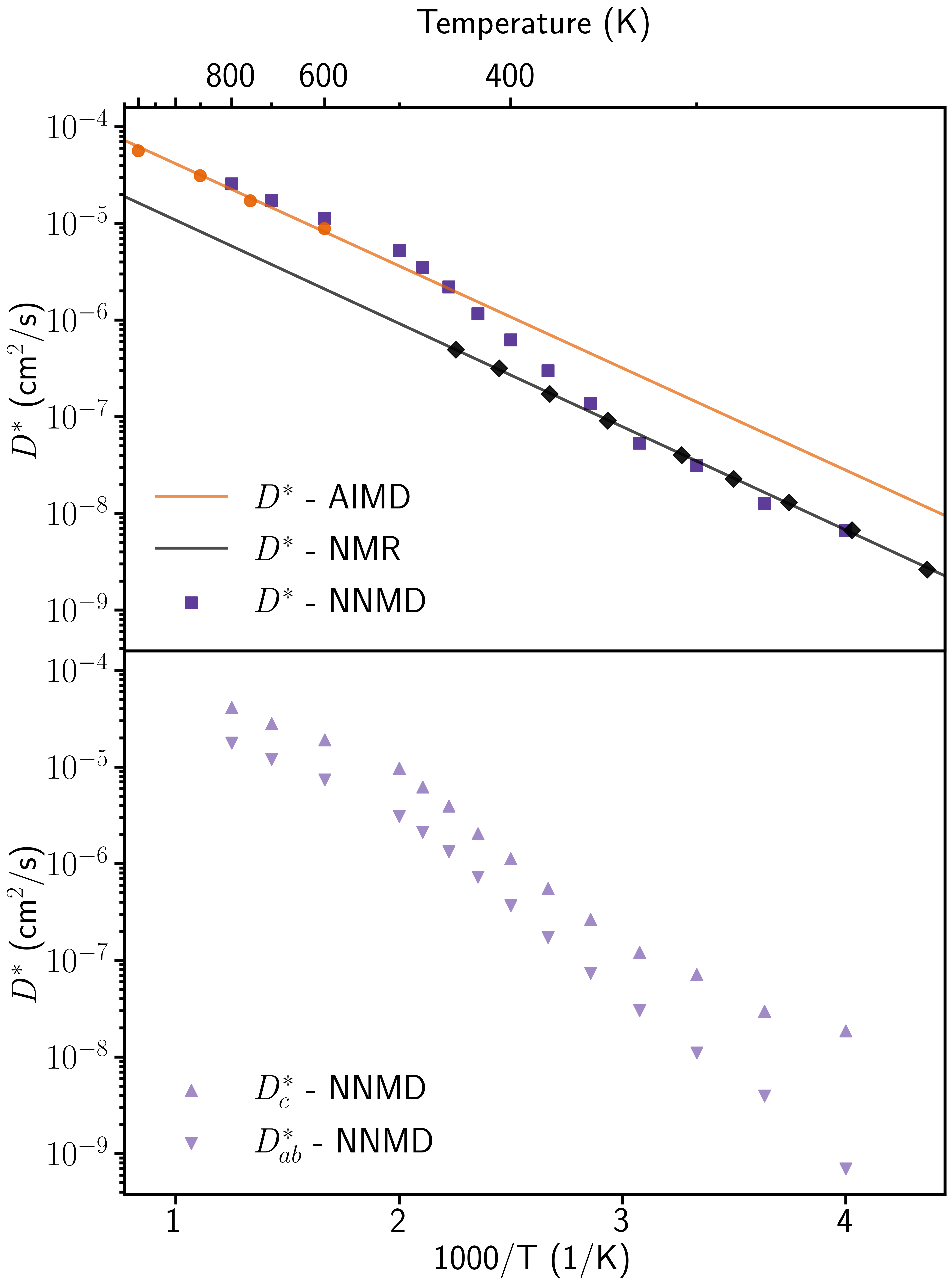}
    \caption{Top is Arrhenius plot of self-diffusivity $D^*$ for Li$_{10}$Ge(PS$_6$)$_2$, including AIMD diffusivity measured by Mo et al. \cite{Mo2012FirstMaterial} and NMR diffusivity measured experimentally by Kuhn et al. \cite{AlexanderKuhn2013TetragonalElectrolytes} Bottom is the Arrhenius plot decomposed by 1D ion conduction along the \textit{c}-channel and 2D ion conduction in the \textit{ab}-plane. Lattice parameters used for this simulation are the 0 K lattice parameters, as relaxed using DFT.}
    \label{fig:arrhenius_plot}
\end{figure}
As seen in the top Arrhenius plot in Figure \ref{fig:arrhenius_plot}, the self-diffusivity $D^*$ values from NNMD match closely with both the AIMD values in the high temperature regime (above 500 K) and with the experimental NMR values in the near-room temperature regime (below 350 K). While the slope of the lines (activation energy) is very similar in the two regimes, in between the two regimes, there is a change in the intercept on the diffusivity axis of nearly an order of magnitude. This can be attributed to a change in the $\Delta S_{migration}$ term included in the pre-factor when plotting $\ln(D^*)$ versus $1/T$. A continuous sublattice phase transition between 350 and 500 K is hypothesized to be responsible for this change in pre-factor, leaving the activation energy (migration barrier) essentially unaltered before and after the transition. From the decomposed \textit{ab}-plane (2D) diffusivity and \textit{c}-channel (1D) diffusivity in the bottom of Figure \ref{fig:arrhenius_plot}, this sublattice phase transition may be attributed to a change in the Li$^+$ sublattice, where the \textit{ab}-plane and \textit{c}-channel diffusivity is drastically reduced below 500 K. Interestingly, below 350 K, the decomposed \textit{ab}-plane diffusivity $D^*_{ab}$ (as denoted by the upside down triangles) continues to follow the same Arrhenius regime as from 350-500 K, while the \textit{c}-channel diffusivity $D^*_{c}$ (as denoted by the triangles) deviates from the Arrhenius behavior established in the 350-500 K regime and remains comparatively high.

We also observed a change in the vibrational density of states that supports the hypothesis of a sublattice phase transition. The vibrational density of states $D(\nu)$, decomposed by ionic species $i$, can be calculated by integrating over the Fourier transform of the velocity $\vec{v}_{i}$ autocorrelation function and weighting by mass $m_i$, as in Equation \ref{eqn:vdos} \cite{Dietschreit2019IdentifyingTransformations,Peters2019CalculatingAssessment}.

\begin{figure}
    \centering
    \includegraphics[height=8cm]{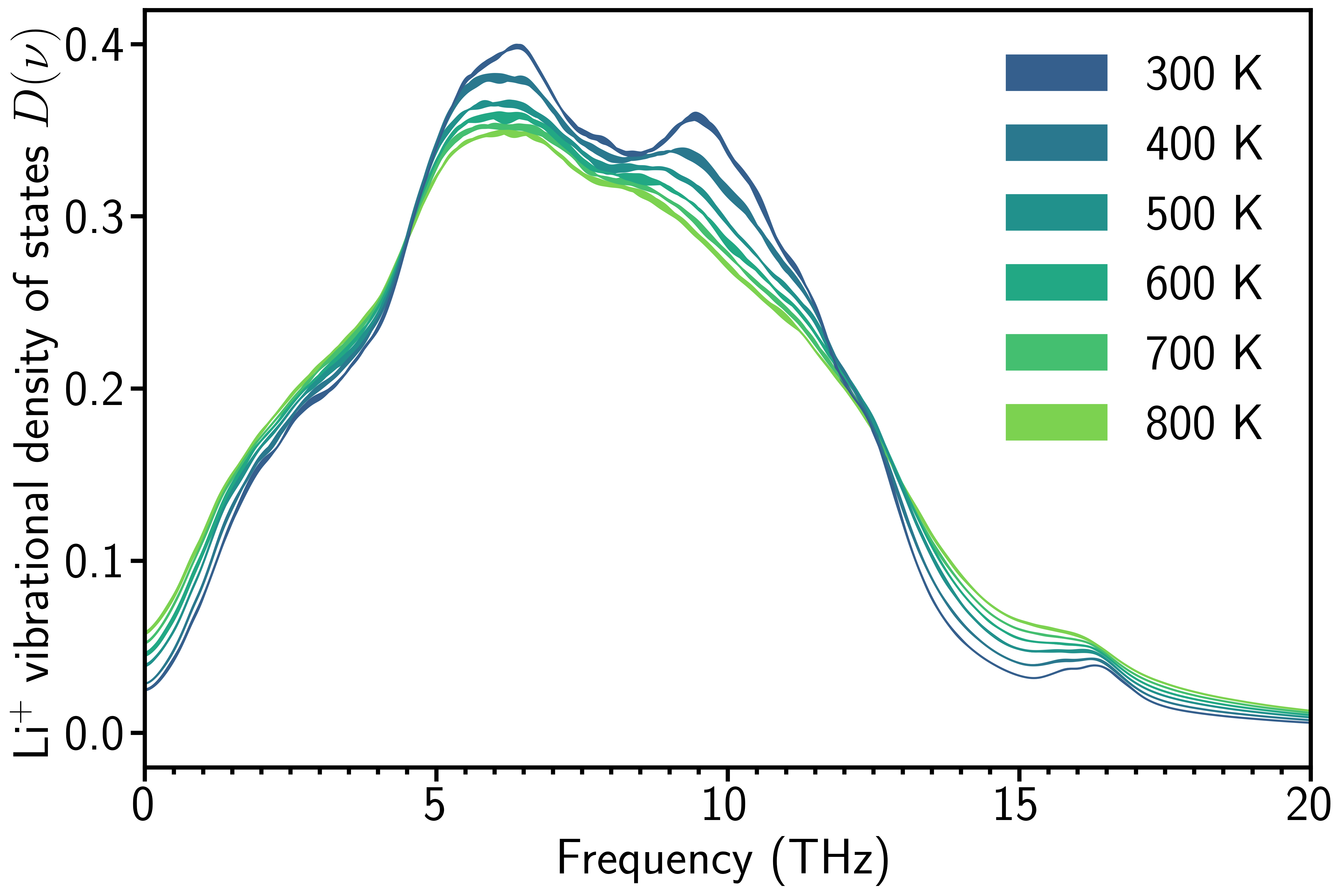}
    \caption{Li$^+$ vibrational density of states spectra, as given by Equation \ref{eqn:vdos}, at each of the temperatures across the sublattice phase transition, from 20 ps of simulation with a 5 fs sampling frequency. Errors depicted by thickness of the density of states curve at a given frequency were obtained from variance over ensemble of MD trajectories, using the same NN potential.}
    \label{fig:vdos}
\end{figure}

\begin{equation}
\label{eqn:vdos}
    D(\nu) = \frac{2}{k_B T} m_i \int < \sum\limits^{N_i}_{i=1}\vec{v}_{i}(\tau) \cdot \vec{v}_{i}(t+\tau) >_{\tau} e^{-i 2 \pi \nu t} dt
\end{equation}

Generally, from the temperature-resolved vibrational density of states spectra in Figure \ref{fig:vdos}, the vibrations in LGPS becomes more harmonic at lower temperature. For reference, the Li dimer harmonic frequency is around 10 THz \cite{Jacox2009VibrationalA}, so the peak in the vibrational density of states just below 10 THz may be attributed to harmonic oscillations of the Li$^+$ sublattice that generally do not contribute to ion conduction. This peak just below 10 THz increases in intensity at lower temperature (300 to 400 K), indicating that Li$^+$ ions' vibrations become more harmonic. The zero-frequency density of states corresponds with anharmonicity and translational motion, which decrease in intensity at lower temperatures. The supplemental information also provides the full vibrational density of states spectra for the other ionic species (see Figure \ref{fig:supporting_vdos} in \ref{sec:supporting_vdos}), which shows that other framework ions such as Ge, P also exhibit more sharply peaked (more harmonic) vibrations at lower temperatures. Notably, there are peaks in the vibrational density of states spectra for P that coincide with the shoulder around 15 THz in the Li$^+$ vibrational density of states. For this reason, this drop in the shoulder around 15 THz with decreasing temperature could be attributed in part to the characteristics of P vibrations, which inhibit Li$^+$ diffusion. 

A larger overlap across a range of frequencies is indicative of stronger coupling and momentum transfer to Li$^+$. Therefore, a broader spectrum for the vibrational density of states for Li$^+$, such as that of LGPS over all the observed temperatures, will generally translate to faster ion conduction. Regardless of the slowdown in the ion conduction mechanism along the \textit{ab}-plane, the vibrational density of states still supports the fact that LGPS is an exceptional ion conductor (even at room temperature). Since integrating over the vibrational density of states is related to vibrational contributions to entropy \cite{McQuarrie2000StatisticalMechanics}, a change in the spectrum for Li$^+$ with respect to temperature may also support a sublattice phase transition, which is generally manifested in a change in the entropy of the system and may translate to a change in the entropy of ion migration pathways.

\begin{figure}
    \centering
    \includegraphics[width=16cm]{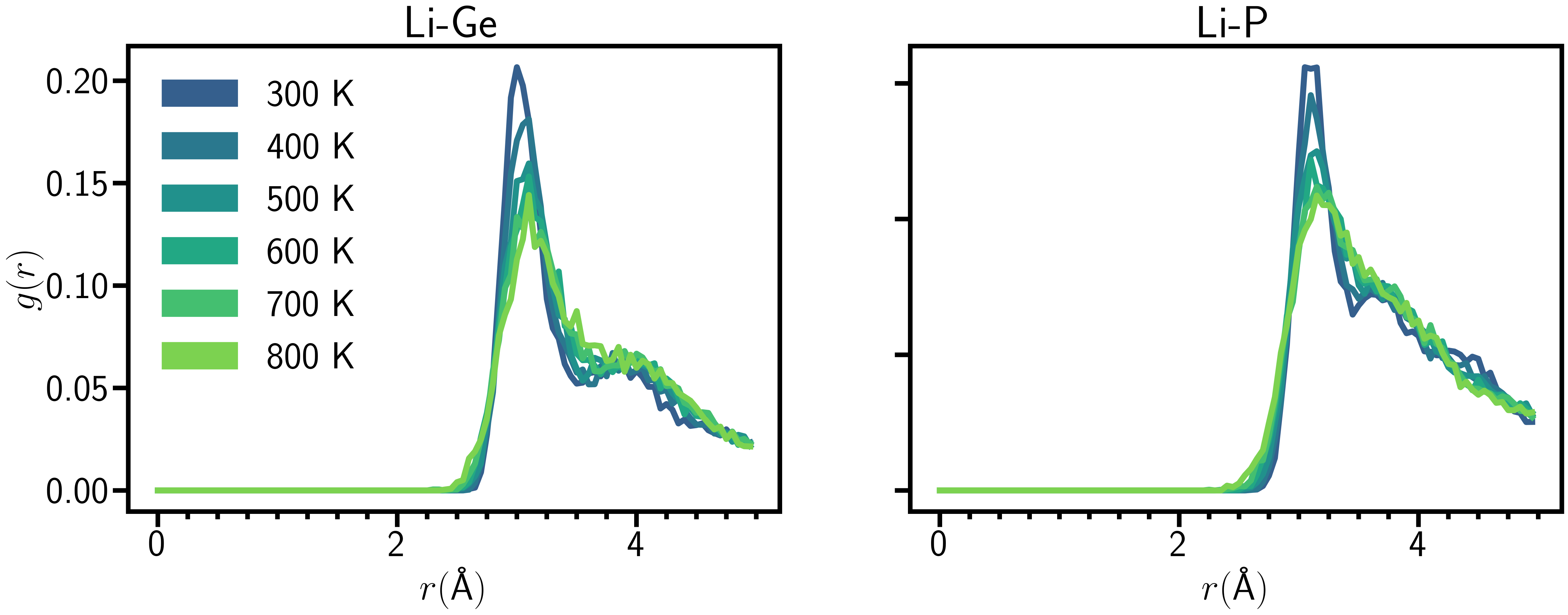}
    \caption{Radial distribution function (RDF), as given by Equation \ref{eqn:rdf}, for Li-P and Li-Ge as a function of temperature from 300 to 800 K. The bin size is 0.05 \AA, and the calculation was taken over an average of 100 frames. Structural changes are evident with the smearing of the two peaks into one peak for Li-P with increasing temperature.}
    \label{fig:rdf}
\end{figure}
    
The radial distribution functions (RDFs), as given by Equation \ref{eqn:rdf} in Figure \ref{fig:rdf} indicate that the structure indeed changes as a function of temperature.
\begin{equation}
    g(\vec{r}) = \frac{1}{\rho} \langle \sum_{i \neq 0} (\delta(\vec{r} - \vec{r_i})) \rangle
    \label{eqn:rdf}
\end{equation}
Specifically, with the Li-P RDF, as the temperature increases from room temperature, two distinct peaks in the RDF at 3.0 \AA and 3.75 \AA, indicative of two inequivalent sites for Li$^+$ network become smeared into one peak at elevated temperature.

\subsection{Correlation in LGPS across temperature regimes}

The self-diffusivity $D^*$, also known as the tracer diffusivity (plotted in Figure \ref{fig:arrhenius_plot}) can be computed by a linear fit of the mean-squared displacement $< \Delta_{self} (t) >$ over time $t$, as shown in Equation \ref{eqn:self_diffusivity}. The mean-squared displacement $< \Delta_{self} (t) >$, or self-correlation, is defined in Equation \ref{eqn:self_correlation}, where $N_c$ is the total number of hopping ions $i$.

\begin{equation}
\label{eqn:self_diffusivity}
     D^* = \frac{1}{6} \lim_{ t \to \infty} \frac{ < \Delta_{self} (t) > }{ t }
\end{equation}

The total ionic conductivity $\sigma_i$ for a solid electrolyte at equilibrium (with no applied potential) depends on the total diffusivity $D_{\sigma}$, which accounts for both the self-diffusivity $D^*$ and the contributions of cross-correlations.

\begin{equation}
\label{eqn:total_diffusivity}
     D_{\sigma} = \frac{1}{6 N_c} < \abs{  \sum^{N_c}_{i=1} [\vec{r}_i(t) - \vec{r}_i(0)] }^2 >
\end{equation}

Thus, the equation for total diffusivity $D$ includes both the self-correlation (Equation \ref{eqn:self_correlation}) and cross-correlation (Equation \ref{eqn:cross_correlation}) contributions embedded in it.
\begin{equation}
\label{eqn:self_correlation}
    \Delta_{self} (t) = \frac{1}{N_c} \sum^{N_c}_{i=1} \abs{\vec{r}_i(t) - \vec{r}_i(0)}^2
\end{equation}
\begin{equation}
\label{eqn:cross_correlation}
    \Delta_{cross} (t) = \frac{1}{N_c} \sum^{N_c}_{i=1} \sum^{N_c}_{i' \neq i=1} \left[ \vec{r}_i(t) - \vec{r}_i(0) \right] \cdot \left[ \vec{r}_{i'}(t) - \vec{r}_{i'}(0) \right]
\end{equation}

In single-ion conductors such as crystalline electrolytes, the cross-correlation $\Delta_{cross} (t)$ would represent the correlation between cations (intra-species correlation) since anions are immobile. An intuitive depiction of the pairwise correlation between different ions is provided in \ref{sec:correlation}. Without computing the cross-correlation directly, the degree of correlation in the system can be assessed by comparing the self-diffusivity $D^*$ with the total diffusivity $D_\sigma$ via the Haven ratio, since the self-correlation and cross-correlation terms sum to the total diffusivity $D_{\sigma}$.

\begin{equation}
\label{eqn:haven_ratio}
    H_r \equiv \frac{ D^* }{ D_{\sigma} }
\end{equation}

The Haven ratio represents the degree to which cross-correlation contributes to the total diffusivity (and thus total ionic conductivity). In a perfectly uncorrelated material system, the Haven ratio would be 1 since the total diffusivity $D_{\sigma}$ is equal to the self-diffusivity $D^*$, as a result of the cross-correlation being negligible.

\begin{figure}
    \centering
    \includegraphics[height=8cm]{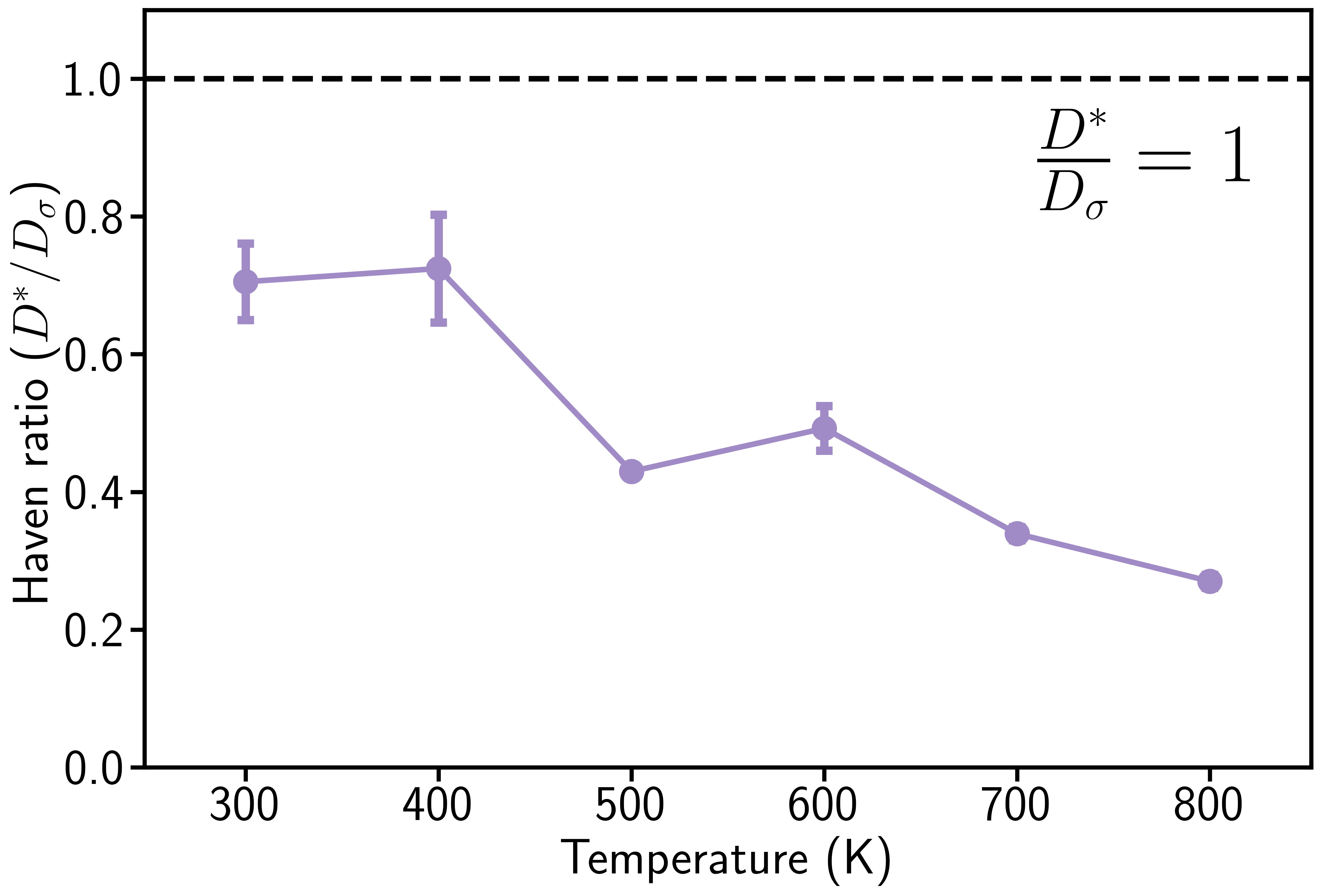}
    \caption{Haven ratio across a range of temperatures, relating the self-diffusivity to the total-diffusivity. A Haven ratio greater than 1 is indicative of destructive cross-correlation, less than 1 is indicative of constructive cross-correlation, and equal to 1 corresponds with perfectly uncorrelated ion motion.}
    \label{fig:haven_ratio}
\end{figure}
While the self-correlation contributes to the self-diffusivity $D^*$ plotted in Figure \ref{fig:haven_ratio}, constructive correlated motion (cross-correlation) dominates at higher temperature, where the electrostatic repulsion of neighboring ions aids in concerted ion hops and boosts total diffusivity relative to the self-diffusivity. However, at low temperatures (in the region of the continuous sublattice phase transition), ion motion is less correlated, with Haven ratios closer to 1 (see Figure \ref{fig:haven_ratio}), consistent with the intuition that electrostatic repulsion from neighboring cations may impede ion motion in a highly concentrated system such as LGPS below a certain temperature.

\begin{figure}
    \centering
    \includegraphics[width=16cm]{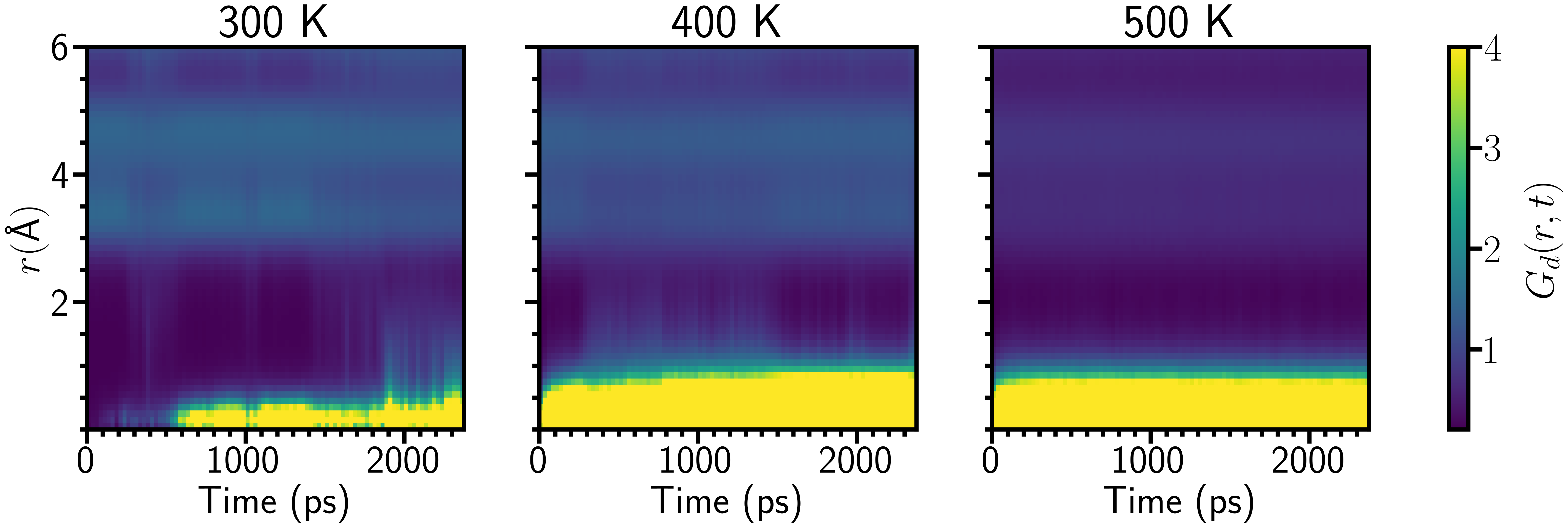}
    \caption{Distinctive van Hove correlation, as given by Equation \ref{eqn:van_hove_correlation}, for Li$^+$ in Li$_{10}$Ge(PS$_6$)$_2$. The bin size was 0.1 \AA, and the sampling frequency was 25 ps. High intensity region approaching $r \rightarrow 0$ are indicative of collective ion motion. The bands appearing above 3 \AA at 300 K indicate that the Li$^+$ sublattice is not liquid-like.}
    \label{fig:van_hove_correlation}
\end{figure}
        
The distinctive van Hove correlation function, as defined in Equation \ref{eqn:van_hove_correlation}, can be interpreted as a time-correlated radial distribution function with respect to an initial reference Li$^+$ ion \cite{Zhu2017AbConductors}. In the following equation,  $\delta()$ is a one-dimensional Dirac delta function,  $r$ is the radial distance from the initial reference Li$^+$ ion, and the sum is performed over all the mobile ions $N_c$, all while being normalized by a factor containing the average number density of the mobile Li$^+$ ions $\rho$.
    \begin{equation}
        G_d (r, t) = \frac{1}{4 \pi r^2 \rho N_c n} \langle \sum_{i \neq j}^{N_c} \delta(r - \abs{\vec{r}_j(t + t_0) - \vec{r}_i(t_0)}) \rangle_{t_0}
        \label{eqn:van_hove_correlation}
    \end{equation}
The peaks (high intensity regions) in the distinctive van Hove correlation near $r=0$ (in the heat maps in Figure \ref{fig:van_hove_correlation}), which are especially evident in the higher temperature cases after several picoseconds, are indicative of collective motion. The high intensity region near $r=0$ describes a neighboring Li$^+$ ion approaching the location of the initial reference Li$^+$ ion after time $t$. This correlation within 1 \AA is likely the result of the electrostatic repulsion of ions playing a role in the concerted ion motion. However, the correlation between cations (cross-correlation) takes 20 times longer to set in at 300 K in comparison to 400 K. Interestingly, in contrast to most correlation effects, which rely on order, this dynamic correlation relies on disorder to aid ion conduction and persists at elevated temperature but is not as evident near room temperature, where part of the Li$^+$ sublattice is more frozen.

The distinctive van Hove correlation function (Figure \ref{fig:van_hove_correlation}) can be read as a RDF by examining one slice in time. From this perspective, it is evident that the Li$^+$ sublattice is more liquid-like at high temperature with broad peaks in the RDF (uniform color in the van Hove correlation plot above 3 \AA). In contrast, at room temperature, the peaks in the RDF are narrower, giving rise to bands appearing around 3 and 5 \AA in the van Hove correlation plot, suggesting the Li$^+$ sublattice cannot be characterized as liquid-like at 300 K.

\subsection{Discussion of discrepancy in sublattice phase transition temperature}

We observed a discrepancy between the self-diffusivity $D^*$ computed by  MD simulations and that measured by NMR \cite{AlexanderKuhn2013TetragonalElectrolytes} in the temperature range between 350 and 450 K, as seen in Figure \ref{fig:arrhenius_plot}. Liang et al. reported a phase transition in the Li$_4$GeS$_4$-Li$_3$PS$_4$ pseudo-binary in the range of 300 K to 400 K, heavily dependent on composition. Using NMR to analyze the local environment around P in the Li-Ge-P-S quaternary system, the phase transition was attributed to changes in the Li sublattice and local environment around P from NMR \cite{Liang2020NewStudy}. The temperature of the phase transition is a very sensitive function of composition. As a result, the discrepancy between NNMD and experiment could be due to a small difference in the composition of LGPS studied experimentally by Kuhn et al. with NMR (in Figure \ref{fig:arrhenius_plot}) and the nominal LGPS structure simulated in this study. Undesired orthorhombic LGPS phase was reported with the high synthesis temperatures by Kuhn et al. \cite{AlexanderKuhn2013TetragonalElectrolytes}. Similarly, there could be impurity phases containing the P$_a$ or P$_c$ instead of the Z arrangement of PS$_4$ and GeS$_4$ polyhedra arrangement simulated in this study. Such impurity phases such as orthorhombic LGPS or $\beta$-Li$_3$PS$_4$, which lower the self-diffusivity of LGPS (relative to the Z tetragonal LGPS simulated in this study) may have resulted in an altered sublattice phase transition temperature. 

In addition to the discrepancy between the self-diffusivity $D^*$ calculated by this study versus experiment, there is also a discrepancy between the self-diffusivity $D^*$ of LGPS calculated by this MD study and other MD studies driven by neural network potentials \cite{Huang2021DeepConductors,Miwa2021MolecularPotential}. At room temperature, both the results of Huang et al. and Miwa et al. overestimate the self-diffusivity $D^*$ as compared to the experimental NMR results (see \ref{sec:supporting_othermlps}).  The discrepancy may arise from either the neural network architectures - the deep potential generation scheme \cite{Zhang2019ActiveSimulation} employed by Huang et al. or the architecture employed by Miwa et al. \cite{Miwa2021MolecularPotential}. Another source of possible disagreement are issues related to convergence of diffusivity statistics with respect to simulation time or finite size effects (discussed in \ref{sec:supporting_finitesize}). The studies of LGPS by Huang et al. \cite{Huang2021DeepConductors} and Miwa et al. \cite{Miwa2021MolecularPotential} use 900-atom and 800-atom supercells, respectively, whereas this study uses a 3200-atom supercell. The present study is the only MD simulation that matched the experimental self-diffusivity $D^*$ (tracer diffusivity) at room temperature, as seen in Figure \ref{fig:arrhenius_plot}. There is a finite size effect reported in LLZO, another highly concentrated solid electrolyte, where smaller supercell sizes overestimate the diffusivity \cite{Klenk2016Finite-sizeLi7La3Zr2O12}, likely because small supercells artificially induce correlation (order). At high temperature this effect is not as evident because of the liquid-like Li$^+$ sublattice, whereas at low temperature it may enforce artificial correlation in the ion hops, resulting in some spurious superionic behavior. In addition, the vibrational modes in a smaller supercell may be different from those in the perfect bulk crystal. Similarly for the LLZO prototype system, there appears to be similar non-Arrhenius behavior approaching room temperature from elevated temperature \cite{Klenk2016Finite-sizeLi7La3Zr2O12,Dai2019ComparisonLi7La3Zr2O12,Qi2021BridgingConductors}. If MD simulations overestimate the diffusivity due to a finite size effect, such important non-Arrhenius behavior may be missed. This non-Arrhenius behavior may be indicative of a sublattice phase transition between two different ion conduction regimes \cite{Dai2019ComparisonLi7La3Zr2O12}, further demonstrating the importance of simulation techniques that are capable of bridging across the varied ion conduction regimes that occur at different temperatures.

\section{Outlook and Conclusion}

Using a fast and accurate NN potential, MD simulations of ionic transport in LGPS have been performed for nearly 40 ns. Two distinct ion conduction regimes have been identified at high temperature and at low temperature, resulting in agreement with historical experimental NMR data and AIMD simulations. A sublattice phase transition is hypothesized to be responsible, and was previously hypothesized but had not been confirmed due to the gap between simulations performed above 500 K and experimental studies performed below 500 K. This sublattice phase transition temperature can be attributed to a drastic drop in the self-diffusivity along the \textit{ab}-plane $D_{ab}^*$. Below this temperature, ion conduction in LGPS can be characterized by a Haven ratio closer to 1, where Li$^+$ ion motion is less correlated and vibrations are more harmonic. Both LLZO and LGPS are unique, in that both materials systems are highly concentrated in Li$^+$, enabling a concerted hopping mechanism due to electrostatic repulsion at elevated temperature. Therefore, a slowdown in ion conduction may be a shared characteristic among other superionic conductors that exhibit a concerted ion hopping mechanism. Below a threshold temperature, the electrostatic repulsion between neighboring mobile ions cannot be overcome by the vibrational modes (distribution of Li$^+$ velocities) available, and consequently ion motion is less concerted at lower temperatures. Despite the change in ion conduction mechanism approaching room temperature from elevated temperature, LGPS remains an exceptional ion conductor at room temperature. More broadly, this study demonstrates the need for fast, accurate methods for driving MD to study ion conduction in crystalline solids at room temperature, without relying on extrapolation.

\section*{Acknowledgments}

The authors thank Simon Axelrod for implementing PaiNN and Dr. Johannes Dietschreit for many fruitful discussions regarding MD simulations and analysis. The authors would also like to acknowledge Dr. Andrey Poletayev for identifying an error in our original analysis of diffusivities. NNMD was run using the computing resources of MIT Supercloud  \cite{Reuther2018InteractiveAnalysis} and Summit at the Oakridge Leadership Computing Facility (through grant number ALCF-MAT237) through the Department of Energy (DOE) Advanced Scientific Computing Research (ASCR) Leadership Computing Challenge (ALCC). DFT work was run with VASP, using the computing resources of Perlmutter and Cori KNL at the National Energy Research Scientific Computing Center (through grant number BES-ERCAP0021172) through the DOE Office of Science's Energy Research Computing Allocations Process. The authors also thank the DOE Advanced Manufacturing Office (grant number DE-EE0009096) and the Toyota Research Institute (TRI) through their Accelerated Materials Design and Discovery (AMDD) program for providing financial support for this work.

\newpage
\section*{Appendix}
\appendix
\setcounter{section}{1}

\subsection{Intuitive description of correlation}\label{sec:correlation}
For the sake of computing the diffusivity by multiplication of matrices or tensors, as can be done when using Equations \ref{eqn:self_correlation} and \ref{eqn:cross_correlation}, the following matrix interpretation of ion displacement and diffusion may be helpful. The displacement can be written as the following vector:
\begin{equation}
    \vec{R}_i (t) = \vec{r}_i (t) - \vec{r}_i (0)
\end{equation}

The self-correlation, which is used to calculate the self-diffusivity (when divided by time), can be intuitively thought of as the on-diagonal elements in the displacement correlation matrix. 

\begin{equation}
    \begin{blockarray}{ccccc}
& Li^+_1 & Li^+_2 &  & Li^+_i \\
\begin{block}{c[cccc]}
  Li^+_1 &  \vec{R}_1 \cdot \vec{R}_1 & {\color{lightgray} \vec{R}_1 \cdot \vec{R}_2 } & {\color{lightgray} \hdots } & {\color{lightgray} \vec{R}_1 \cdot \vec{R}_i } \\
  Li^+_2 & {\color{lightgray} \vec{R}_2 \cdot \vec{R}_1 } & \vec{R}_2 \cdot \vec{R}_2 & {\color{lightgray} \hdots } & {\color{lightgray} \vec{R}_2 \cdot \vec{R}_i } \\
  & {\color{lightgray} \vdots } & {\color{lightgray} \vdots } & \ddots & {\color{lightgray} \vdots} \\
  Li^+_i & {\color{lightgray} \vec{R}_i \cdot \vec{R}_1 } & {\color{lightgray} \vec{R}_2 \cdot \vec{R}_2 } & {\color{lightgray} \hdots } & \vec{R}_i \cdot \vec{R}_i \\
\end{block}
\end{blockarray}
\end{equation}

The cross-correlation, which is added to the self-correlation to calculate the total diffusivity, can be intuitively thought of as the off-diagonal elements in the displacement correlation matrix.

\begin{equation}
\begin{blockarray}{ccccc}
& Li^+_1 & Li^+_2 &  & Li^+_i \\
\begin{block}{c[cccc]}
  Li^+_1 & {\color{lightgray} \vec{R}_1 \cdot \vec{R}_1 } & \vec{R}_1 \cdot \vec{R}_2 & \hdots & \vec{R}_1 \cdot \vec{R}_i \\
  Li^+_2 & \vec{R}_2 \cdot \vec{R}_1 & {\color{lightgray} \vec{R}_2 \cdot \vec{R}_2 } & \hdots & \vec{R}_2 \cdot \vec{R}_i \\
  & \vdots & \vdots & {\color{lightgray} \ddots } & \vdots \\
  Li^+_i & \vec{R}_i \cdot \vec{R}_1 & \vec{R}_2 \cdot \vec{R}_2 & \hdots & {\color{lightgray} \vec{R}_i \cdot \vec{R}_i } \\
\end{block}
\end{blockarray}
\end{equation}

\subsection{Additional calculation details for VASP DFT}\label{sec:supporting_dftdetails}
The \textit{ab initio} calculations were carried out as follows: an energy cutoff of 520 eV was used for the plane-wave basis set. The Perdew-Burke-Ernzerhof (PBE) generalized gradient approximation (GGA) functional was used \cite{Perdew1996GeneralizedSimple}, but a functional with a higher level of theory could have been used for training data calculations, embedding into the (NN) potential a higher accuracy, if desired. A 2 × 2 × 2 $\Gamma$-centered k-point mesh was used to sample the Brillouin zone. The energy criterion for self-consistency was when the energy difference between subsequent electronic steps was less than 1 $\mu$eV. For NEB and structure optimizations (relaxations), the convergence criterion was set to an atomic force tolerance of 0.01 eV/\AA.

\subsection{Data and model availability}\label{sec:supporting_data}

The dataset used for training models and the trained PyTorch models are available on \href{https://figshare.com/projects/Dataset_and_machine-learning_potentials_for_LGPS_solid_electrolyte/152601}{figshare here}. As noted in the main body of the paper, each datum of this dataset contains the DFT-computed forces and energy corresponding with each structure (coordinates and lattice). Currently, the PaiNN model can be used to run MD (with an ASE interface) through the  \href{https://github.com/learningmatter-mit/NeuralForceField}{NeuralForceField} GitHub repository, which is a general framework that can be used for running MD with NN potentials, while the NequIP model can be run using either an ASE interface or LAMMPS interface with the patch provided through the \href{https://github.com/mir-group/nequip}{NequIP} GitHub repository \cite{Batzner2022E3-EquivariantPotentials}. The raw MD trajectories are available upon request.

\subsection{Full vibrational density of states for all ions across temperatures}\label{sec:supporting_vdos}

\begin{figure}[H]
    \centering
    \includegraphics[width=15cm]{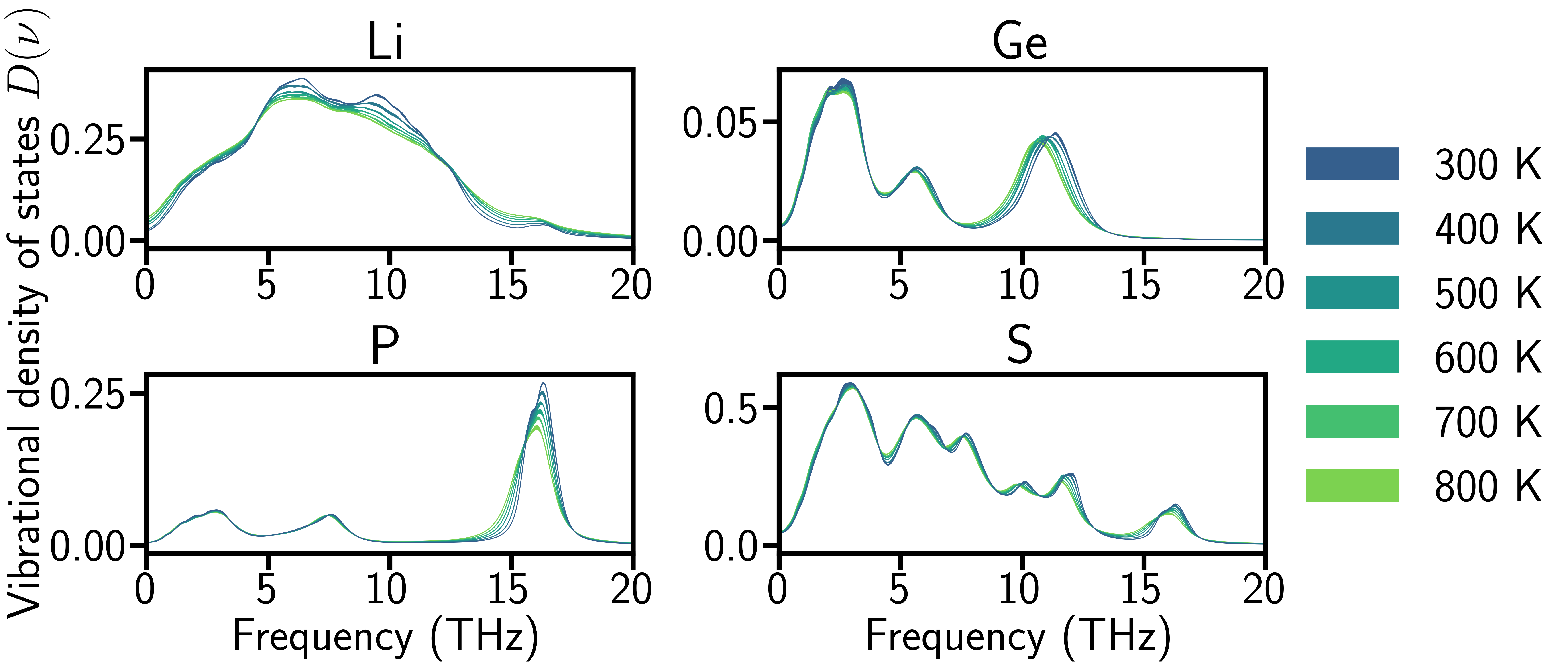}
    \caption{Vibrational density of states spectra decomposed by element, as given by Equation \ref{eqn:vdos}, at each of the temperatures across the sublattice phase transition, from 20 ps of simulation with a 5 fs sampling frequency. Select peaks at specific frequencies for Ge, P, S become more sharply peaked (higher intensity) at lower temperatures as compared to higher temperatures, indicative of greater harmonicity at low temperatures}
    \label{fig:supporting_vdos}
\end{figure}

In Figure \ref{fig:supporting_vdos}, the peaks corresponding to the vibrational modes of the framework ions Ge, P, and S in general shift to lower frequencies with increasing temperature. Notably, the peaks just above 15 THz for P and S broaden at elevated temperature and smear out the shoulder that is more distinctly present at room temperature. Such peak broadening can be attributed to less harmonic vibrations at elevated temperatures since harmonic vibrations would, in comparison, appear as narrower delta functions.

\newpage
\subsection{Thermostat parameters and their effect on diffusivity}

\begin{figure}[H]
    \centering
    \includegraphics[width=15cm]{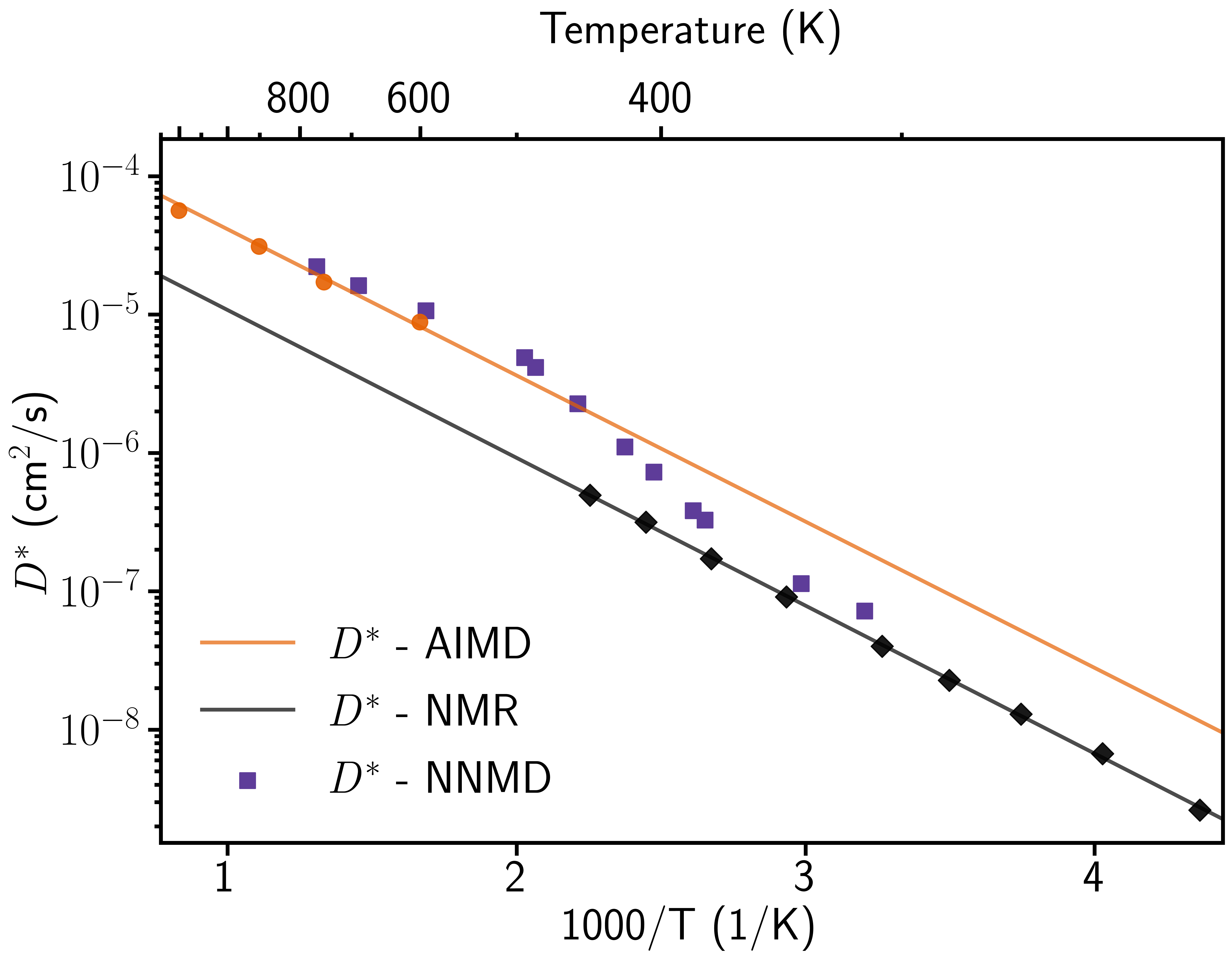}
    \caption{Arrhenius plot of self-diffusivity $D^*$ for Li$_{10}$Ge(PS$_6$)$_2$, including AIMD diffusivity measured by Mo et al. \cite{Mo2012FirstMaterial} and NMR diffusivity measured experimentally by Kuhn et al. \cite{AlexanderKuhn2013TetragonalElectrolytes} This is MD run without a thermostat (i.e. NVE ensemble), where lattice parameters used for this simulation are the 0 K lattice parameters.}
    \label{fig:arrhenius_nve}
\end{figure}

The Nos\'e-Hoover ttime parameter as implemented in ASE for the Arrhenius plot in Figure \ref{fig:arrhenius_plot} was set to 20, where the timestep was set to 1 fs. With the NVE ensemble (no thermostat), the supercell thermally equilibrated well, and the sublattice phase transition is still evident in Figure \ref{fig:arrhenius_nve}, demonstrating that the thermostat does not interfere with the ion hopping dynamics or the phenomena investigated here. Each of these NNMD runs lasted 10 ns, which is why convergence may have not been reached for some of the lower temperatures in Figure \ref{fig:arrhenius_nve}.
The spacing between temperatures was also not consistent since the only parameter that can strictly be controlled with NVE is the initialization temperature, whereas the average temperature is subject to fluctuations.

\newpage
\subsection{Accounting for thermal expansion and its effect on diffusivity}

\begin{figure}[H]
    \centering
    \includegraphics[width=15cm]{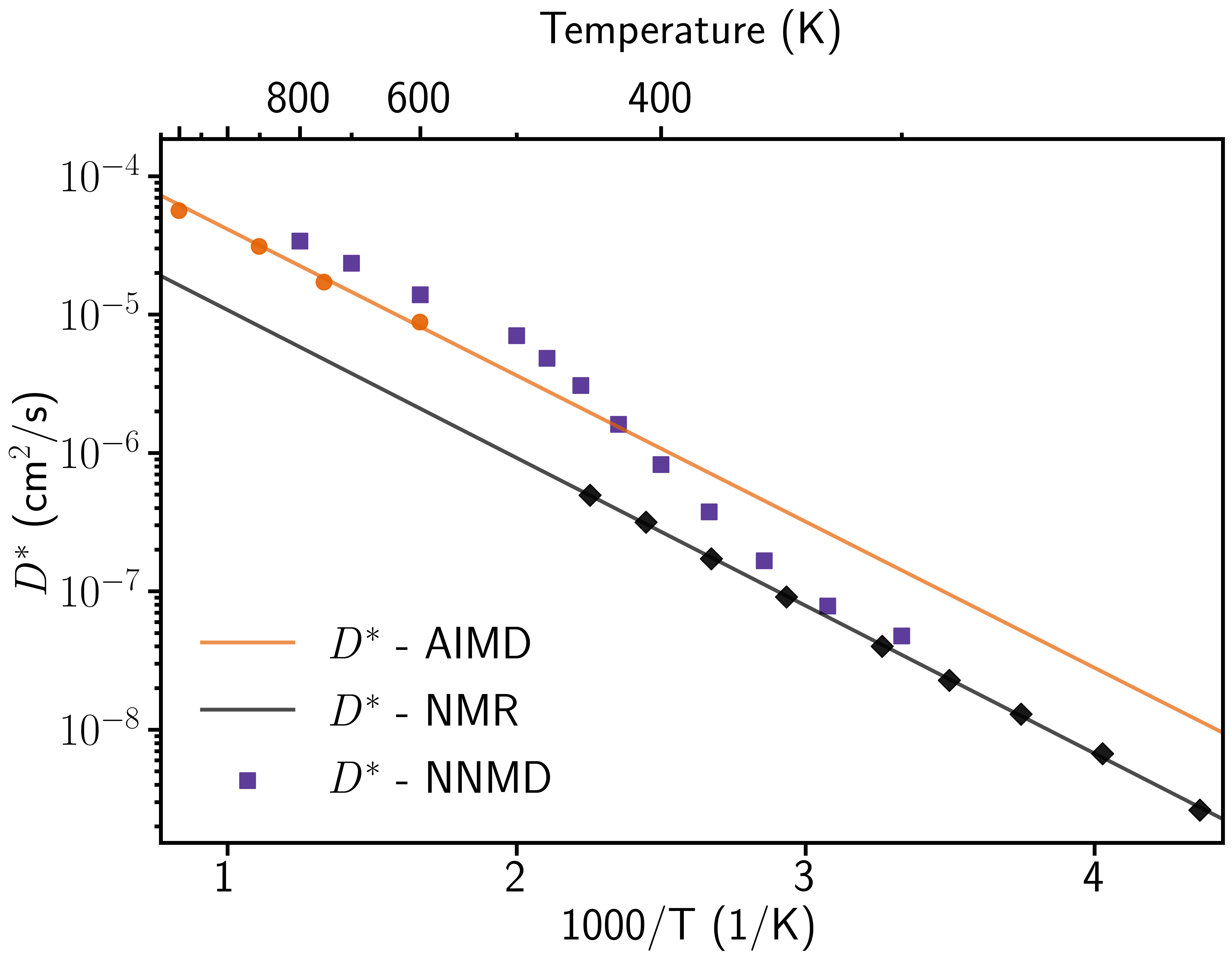}
    \caption{Arrhenius plot of self-diffusivity $D^*$ for Li$_{10}$Ge(PS$_6$)$_2$, including AIMD diffusivity measured by Mo et al. \cite{Mo2012FirstMaterial} and NMR diffusivity measured experimentally by Kuhn et al. \cite{AlexanderKuhn2013TetragonalElectrolytes} This is MD run with the Nos\'e-Hoover thermostat (i.e. NVT ensemble). Lattice parameters used for this are the experimental lattice parameters that correspond with each temperature on the Arrhenius plot, accounting for thermal expansion.}
    \label{fig:arrhenius_exptlattparams}
\end{figure}

Since 0 K lattice parameters are non-physical, the thermal expansion of the lattice should be accounted for, using the experimental lattice parameters from elevated temperature to demonstrate that the sublattice phase transition still occurred. The results from this are shown in Figure \ref{fig:arrhenius_exptlattparams}. Each of these NNMD runs lasted 10 ns, which is why convergence may have not been reached for some of the lower temperatures. Indeed, at low temperature, the diffusivities in Figure \ref{fig:arrhenius_exptlattparams} are generally about the same as in Figure \ref{fig:arrhenius_plot} since the lattice parameters at low temperature are close to the lattice parameters at 0 K. In general there is poorer agreement between NNMD and AIMD at higher temperature, but this is to be expected, given that the reference AIMD by Mo et al. \cite{Mo2012FirstMaterial} was run using 0 K lattice parameters.

\newpage
\subsection{Effect of finite simulation size on diffusivity}\label{sec:supporting_finitesize}

\begin{figure}[H]
    \centering
    \includegraphics[width=15cm]{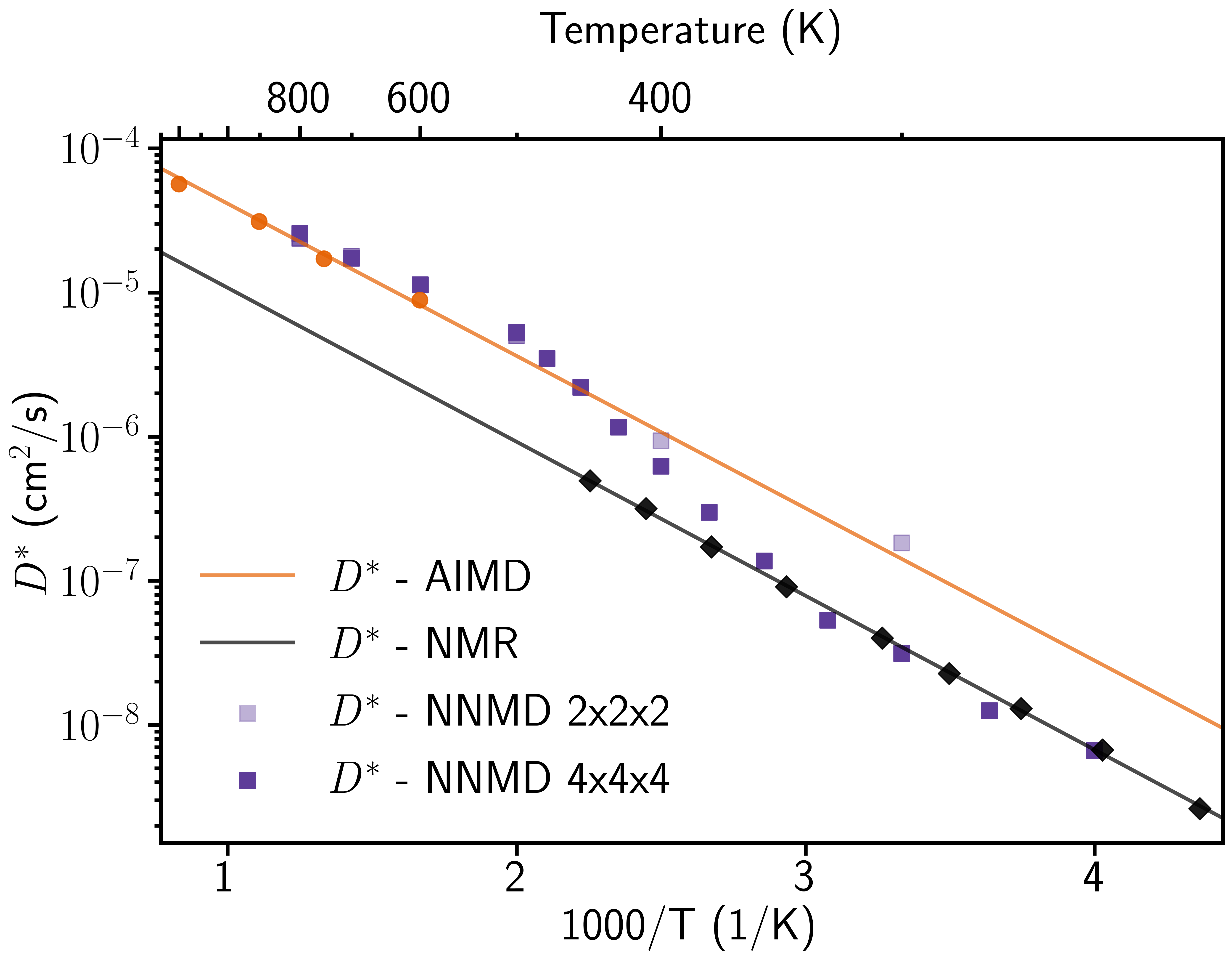}
    \caption{Arrhenius plot of self-diffusivity $D^*$ for Li$_{10}$Ge(PS$_6$)$_2$, including AIMD diffusivity measured by Mo et al. \cite{Mo2012FirstMaterial} and NMR diffusivity measured experimentally by Kuhn et al. \cite{AlexanderKuhn2013TetragonalElectrolytes} This is MD run with the Nos\'e-Hoover thermostat (i.e. NVT ensemble) and varied supercell sizes (2x2x2 = 400 atoms, 4x4x4 = 3200 atoms). The lattice parameters used for this simulation are the 0 K lattice parameters.}
    \label{fig:arrhenius_finitesize}
\end{figure}

As seen in Figure \ref{fig:arrhenius_finitesize}, there is good agreement in diffusivity values at temperatures above 500 K across all supercell sizes. However, below 500 K, which is the region of interest for the sublattice phase transition, the smaller 2x2x2 supercell (400 atoms), overestimates the diffusivity. Similarly, as seen in the next section \ref{sec:supporting_othermlps}, comparing our work and Huang et al.'s work on MD simulations with a 900-atom supercell using a NN potential trained on the same underlying DFT data (PBE GGA) \cite{Huang2021DeepConductors}, the diffusivity is overestimated at lower temperatures when a small supercell is used. In both of these instances, this may be due to a finite size effect, which Huang et al. also observed \cite{Huang2021DeepConductors}. Such a finite size effect could be attributed to smaller supercell sizes, artificially inducing correlation (order) in comparison to a true infinite bulk crystal. For this reason, we opted for a larger 4x4x4 supercell (3200 atoms) for production MD simulations, which would be able to capture the phenomena of interest \textemdash a change in the ion conduction mechanism.

\subsection{Comparison with other machine learning potentials for LGPS}\label{sec:supporting_othermlps}

\begin{figure}[H]
    \centering
    \includegraphics[width=15cm]{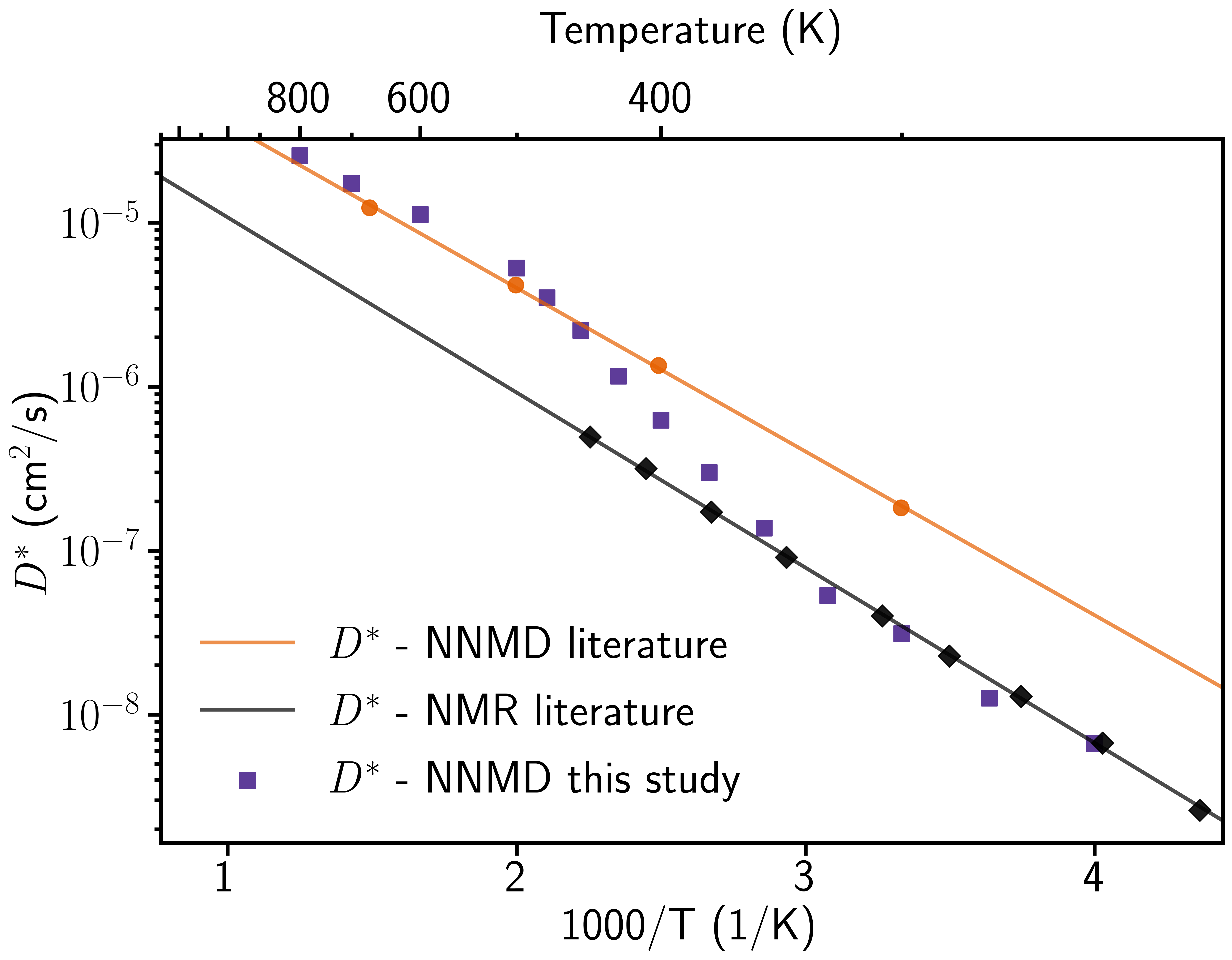}
    \caption{Arrhenius plot of self-diffusivity $D^*$ for Li$_{10}$Ge(PS$_6$)$_2$, including NNMD literature values measured by Huang et al. \cite{Huang2021DeepConductors} and NMR literature values measured experimentally by Kuhn et al. \cite{AlexanderKuhn2013TetragonalElectrolytes} This is MD run with the Nos\'e-Hoover thermostat (i.e. NVT ensemble). The supercell size used in our NNMD simulation was 4x4x4 (3200 atoms), while the supercell size used by NNMD from literature (Huang et al.) was 900 atoms. The lattice parameters used for this simulation are the 0 K lattice parameters.}
    \label{fig:arrhenius_discrepancy}
\end{figure}

As seen in Figure \ref{fig:arrhenius_discrepancy}, there is a discrepancy between the diffusivity calculated by NNMD simulations previously reported by Huang et al. \cite{Huang2021DeepConductors} and experiment \cite{AlexanderKuhn2013TetragonalElectrolytes} approaching room temperature. This may be attributed to the finite size effect discussed in the previous section \ref{sec:supporting_finitesize}.

\newpage
\section*{References}

\bibliographystyle{iopart-num}
\bibliography{references}


\end{document}